\documentclass{article}

\usepackage{PRIMEarxiv}

\usepackage[utf8]{inputenc} % allow utf-8 input
\usepackage[T1]{fontenc}    % use 8-bit T1 fonts
\usepackage{hyperref}       % hyperlinks
\usepackage{url}            % simple URL typesetting
\usepackage{booktabs}       % professional-quality tables
\usepackage{amsfonts}       % blackboard math symbols
\usepackage{nicefrac}       % compact symbols for 1/2, etc.
\usepackage{microtype}      % microtypography
\usepackage{lipsum}
\usepackage{fancyhdr}       % header
\usepackage{graphicx}       % graphics
\graphicspath{{media/}}     % organize your images and other figures under media/ folder

\title{ Unsupervised Image to Image translation for multiple retinal pathology synthesis in optical coherence tomography scans
}

\author{
  Hemanth Pasupuleti\thanks{\textit{\underline{Corresponding Author}}: satyasaihemanth.p@gmail.com}\\
Department of Computer Science and Engineering \\
Indian Institute of Information Technology, Sri City, India. \\
  \texttt{satyasaihemanth.p@gmail.com; satyasaihemanth.p18@iiits.in} \\
   \And
  G. N. Girish \\
  Department of Computer Science and Engineering \\
Indian Institute of Information Technology, Sri City, India. \\
  \texttt{girishanit@gmail.com; girish.gn@iiits.in} \\
}

\begin{document}
\maketitle

\begin{abstract}
Image to Image Translation (I2I) is a challenging computer vision problem used in numerous domains for multiple tasks. Recently, ophthalmology became one of the major fields where the application of I2I is increasing rapidly. One such application is the generation of synthetic retinal optical coherence tomographic (OCT) scans. Existing I2I methods require training of multiple models to translate images from normal scans to a specific pathology: limiting the use of these models due to their complexity. To address this issue, we propose an unsupervised multi-domain I2I network with pre-trained style encoder that translates retinal OCT images in one domain to multiple domains. We assume that the image splits into domain-invariant content and domain-specific style codes, and pre-train these style codes. The performed experiments show that the proposed model outperforms state-of-the-art models like MUNIT and CycleGAN synthesizing diverse pathological scans.
\end{abstract}

% keywords can be removed
\keywords{Optical Coherence Tomography, Generative Adversarial Networks, Image to Image translation, Retinal Disorders,  \and Retinal Pathology}

\section{Introduction}
The human eye is one of the vital and complex organs which provides the ability to see and perceive the surrounding world. When the light enters the pupil and strikes the retina, it is converted into nerve signals that are processed by the brain. Due to the aging population and increase in the prevalence of diabetes, diseases like age-related macular degeneration(AMD), and Diabetic Macular Edema(DME) became reasons for the majority of vision loss \cite{pmid22559899,pmid22585044}. 

Optical Coherence Tomography (OCT) is a leading noninvasive imaging technique utilized to acquire cross-sectional retinal imaging in ophthalmology\cite{pmid1957169}. It helps doctors to diagnose diseases, monitor their progress, and navigate during surgery. Thus, playing a vital role in the treatment of retinal diseases. Several deep learning methods were employed to automate this process and tackle various image analysis tasks like detection and segmentation in OCT imaging \cite{girish2019depthwise, girish2018segmentation, SCHLEGL2018549,pmid33329940}. However, to achieve these tasks, a large dataset is usually required.

Traditional image augmentation methods like image shifting, rotation, scaling, and deformation are used widely in medical imaging but limit the diversity of the features obtained from the augmented images \cite{girish2018segmentation}. Goodfellow \textit{et al} proposed Generative Adversarial Networks (GANs) \cite{NIPS2014_5ca3e9b1} that led to the emergence of using synthetically generated data for improving the performance of various medical image analysis tasks with deep learning \cite{FRIDADAR2018321,8363678}.

To generate images of the desired pathology from Normal B-Scan images, we propose a Generative Adversarial Network (GAN) model in this paper. We evaluate our model with other existing models by generating both prevalent and rare diseases.

\section{Related Work}
\label{sec:headings}

%\textbf{Image to Image translation.} 
The advent of GANs led to their application in various fields like image generation\cite{2015arXiv151106434R}, super-resolution\cite{Bulat_2018_ECCV}, image inpainting\cite{2018arXiv180307422D,Yu_2018_CVPR}, etc.. They usually contain two networks: a generator that learns to generate images and a discriminator that distinguishes between the generated fake image and real image. Conditional GAN (cGAN) \cite{2014arXiv1411.1784M} is a variant of GAN where the class knowledge is provided into the network to impose control on the generated image. Image to Image Translation (I2I) falls into one of the cGAN applications where the model learns mapping to translate input images between different domains. Initially, researchers used input-output pair images to achieve the I2I task between two domains \cite{8100115}. However, obtaining these paired images is often difficult for many tasks and CycleGAN \cite{8237506} alleviates this problem by using unpaired images. CycleGAN displayed that it can produce high-quality images but it lacks in the diversity that is addressed by MUNIT \cite{HUANG2018}.

% O N Hassan \textit{et al} \cite{2020arXiv201004552H} used over 25,000 OCT images for training conditional GAN to predict the possible transition of glaucoma disease by reconstructing with the previous measurements. Liu Y \textit{et al} \cite{Liu1735} used pix2pixHD to generate the response to a pathology imaging for a treatment. Jeihouni \textit{et al} \cite{9506291} brought together segmentation and image super resolution to generate high resolution segmentation of retinal layers from low resolution input images. Mahapatra \textit{et al} \cite{Mahapatra_2020_CVPR} proposed a new model that can augment images without losing the geometrical relationship by utilizing the segmentation mask.

Recently, Zheng \textit{et al} assessed the quality of high-resolution retinal OCT images generated by GANs \cite{pmid32832202}. The generated images were evaluated by two ophthalmologists and it was determined that synthetic retinal OCT scans aid in training and educational activities and can also serve as data augmentation to enhance the existing dataset for building machine learning models. Xiao \textit{et al} \cite{9098320} proposed an open set recognition system by using synthetic OCT images. These generated images are considered to be of unknown class and thus making the classifier able to detect rare or unknown diseases. Furthermore, a study was conducted focusing on the role of GAN-generated images in improving the accuracy of classifiers for detecting rare diseases \cite{pmid33492598}. They trained 5 CycleGAN models where each CycleGAN model translates from a normal retinal OCT image to one rare disease. The translated images were then evaluated by experts and they also show that these synthetic images help increase the accuracy of the classifier. 

In all of these works, even though GANs have shown promising results they lack control between different classes since the models that were used only translate between two domains. Due to this, if we want to translate normal images into pathological images then we have to train an individual model for each pathology thus limiting the application of GANs in retinal imaging as it requires a lot of time. Models like StarGANv2\cite{9157662} learn a many-to-many mapping between multiple domains which is not necessary since we only have to translate from a normal image to multiple pathological images. Hence, in this work, we propose a model that can generate multiple pathological images from normal images. Inspired by StarGANv2, we adapt the one generator and one discriminator policy while training the model in an unpaired fashion like MUNIT, without showing the real pathological images to the generator.

% See Section \ref{sec:headings}.

% \subsection{Headings: second level}
% \lipsum[5]
% \begin{equation}
% \xi _{ij}(t)=P(x_{t}=i,x_{t+1}=j|y,v,w;\theta)= {\frac {\alpha _{i}(t)a^{w_t}_{ij}\beta _{j}(t+1)b^{v_{t+1}}_{j}(y_{t+1})}{\sum _{i=1}^{N} \sum _{j=1}^{N} \alpha _{i}(t)a^{w_t}_{ij}\beta _{j}(t+1)b^{v_{t+1}}_{j}(y_{t+1})}}
% \end{equation}

% \subsubsection{Headings: third level}
% \lipsum[6]

% \paragraph{Paragraph}
% \lipsum[7]

\section{Approach}
\label{sec:others}
In this section, we discuss the proposed method to generate multi-domain retinal OCT images.

\subsection{Framework}
\begin{figure}
  \centering
  \includegraphics[scale=0.4]{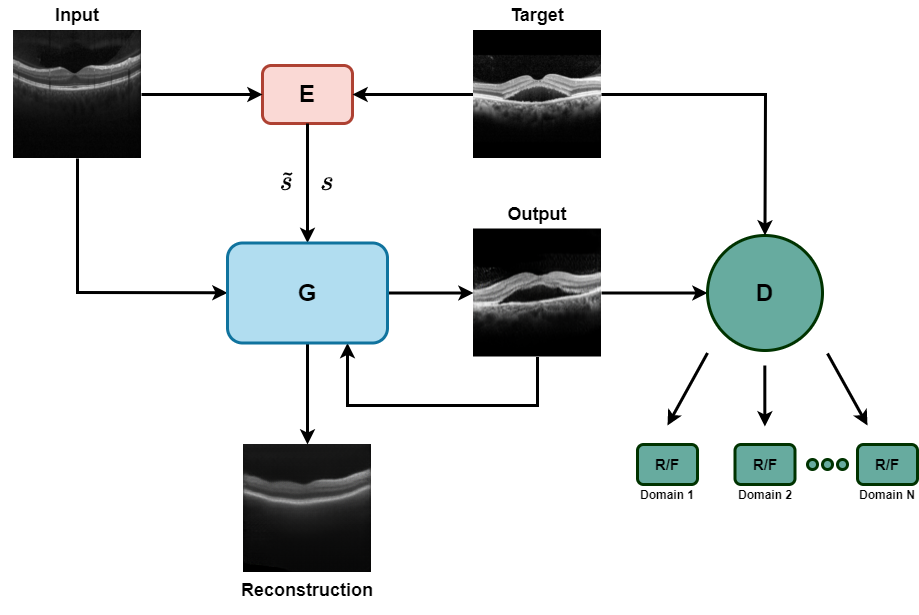}
  \caption{\textbf{Proposed Framework.} The network architecture mainly consists of 3 parts: \textbf{(a)} Pre-trained Style Encoder: We have one style encoder ${E}$ that extracts the style codes of respective domains. These codes are then used for reference-guided synthesis. Here ${s,\widetilde{s}}$ represent the corresponding style codes for target and input images. \textbf{(b)} Generator: We also use only one generator ${G}$ that translates the input image into multiple target domains by utilizing the style code. \textbf{(c)} Discriminator: The discriminator ${D}$ captures the relation between the real and fake images. It has multiple branches with outputs specific to the input domain that can be selected while training.}
  \label{fig:fig1}
\end{figure}

Consider we have images that are normal without any pathology in the domain ${X}$ and all the target pathological images of different classes be ${Y_1, Y_2, Y_3,..., Y_n}$ (where ${n}$ represents the class). Our goal is to learn the mapping ${X\rightarrow\left\{Y_n\mid n>0\right\}}$ to generate the target image. Figure \ref{fig:fig1} represents the proposed architecture for unsupervised multi-domain I2I translation of OCT images.
 
% There are many methods that require Z as the condition to generate images. Recently Style embeddings were used to translate images between multiple domains [references here].

% However, in multiple domains scenario there is either a need to train N models to learn the mapping from one to N domains or the generator is required to feed the target domains. 

% In this approach, we adapt the unified network strategy of having only one generator and one discriminator like stargan and starganv2 while training the model in an unsupervised approach like Cyclegan,munit,drit etc... That is, we translate images from one to many domains without showing the target images to the generator rather than solely depending on the output of the discriminator. 
\textbf{Style Encoder Pre-training:} Gram matrices have been introduced to represent the stylistic features of a reference image in neural style transfer \cite{2015arXiv150806576G}. Many models use learned style encoding that is similar to the style encoding obtained from gram matrices to enforce condition on generated image \cite{HUANG2018,8953766}. The main problem with this approach is that they depend on the target dataset and don't capture styles that are not well represented. Recently, Meshry \textit{et al} \cite{9578574} showed that style encoder pre-training mitigates this issue due to a more robust latent space representation and produces expressive results. This encoder pre-training helps us to gain control over the pathologies especially in retinal OCT imaging where the diseases may have overlapping characteristics. It also enables us to generate various retinal OCT images with desired characteristics. 

We aim to train the style embedding network such that the output embeddings of similar classes are close together. While Meshry \textit{et al} \cite{9578574} used triplet loss to train the network by selecting triplets using style distance metric, there has been a series of work exploring different triplet mining techniques and losses \cite{7298682,Harwood_2017_ICCV,Nina_2019_ICCV,9093432}. In this work, we proceed to train the style encoder by using Easy Positive Hard Negative triplet mining proposed by Xuan \textit{et al} \cite{9093432}.

\textbf{Style Encoder:} Given an input image ${x}$, our style encoder ${E}$, produces the style embeddings ${z = E(x)}$ that are lower-dimensional projections of Gram matrices.

\textbf{Generator:} Providing an image ${x}$, our generator ${G}$ generates the output image ${G(x,s)}$ translating input to the target domain. Here, ${s}$ is the style code of the target domain that is obtained from the pre-trained style encoder ${E}$. We feed the style information into the generator by using Adaptive Instance Normalization (AdaIN)\cite{8953766}.

\textbf{Training:} There are two stages of training in the proposed approach:
\begin{itemize}
\item \textbf{Stage 1:} For all the classes present in a given dataset, cluster the style embeddings produced by the style encoder ${E}$.
\item \textbf{Stage 2:} After successful training, we freeze the weights of ${E}$ and train the generator ${G}$. The style encoder ${E}$ delivers the target domain information into the generator which translates the input image.
\end{itemize}

Following the works of Meshry \textit{et al}, an additional stage can be employed that requires combined fine-tuning of both style encoder and generator. Here, we unfreeze the style encoder weights and train the entire model as a single-stage architecture. However, in this work, we only present the results obtained by training with the earlier two stages and make this stage completely optional since careful hyperparameter tuning is required.

% The discriminator is a relativistic multi task discriminator. The discriminator D(x,y) takes input x and label y to output multiple branches representing multiple labels. To further make the training more stable we employ spectral normalization in the discriminator and R1 Regularization.
\textbf{Discriminator:} Recently relativistic discriminators\cite{2018arXiv180700734J} have proved their ability to produce high quality images in various domains\cite{Wang_2018_ECCV_Workshops,Du2021,Liu_2019_ICCV}. We adopt this relative discriminator with the multi-task discriminator\cite{9010865} to design a relativistic multi-task discriminator. Given an image ${z}$ and label ${\widehat{z}}$, the discriminator ${D}$ outputs multiple branches with each branch representing an individual label. For each branch, the outputs range from 0 to 1 representing fake and real images. The discriminator only targets to optimize the branch corresponding to the given label establishing an intraclass relationship. To further make the training more stable, we employ spectral normalization\cite{2018arXiv180205957M} and R1 regularization\cite{2018arXiv180104406M}.

\subsection{Losses}

\textbf{Adversarial Loss:} There are a variety of losses with their functionality proposed for GANs. To avoid bad basins that result in the mode collapse we use Relative pairing hinge loss \cite{NEURIPS2020_738a6457} for stable training and faster convergence. 

\begin{equation}
    L_{D}^{adv} = \mathbb{E}_{x,y}\left[max(0,1+(D_{\widehat{y}}(G(x,s))-D_{\widehat{y}}(y)))\right]
\end{equation}

\begin{equation}
    L_{G}^{adv} = \mathbb{E}_{x,y}\left[max(0,1+(D_{\widehat{y}}(y)-D_{\widehat{y}}(G(x,s))))\right]
\end{equation}

where ${\widehat{y}}$ is the corresponding class label for the target domain ${y}$ and ${D_{\widehat{y}}(.)}$ denotes the output of the discriminator for the target class label ${\widehat{y}}$. The ${s = E(y)}$ is the style embedding generated for the reference image ${y}$.

\textbf{Cycle Consistency loss:} To make sure that the model is preserving the source characteristics, we use cycle consistency loss. After generating image ${G(x,s)}$ from image ${x}$ we again try to reconstruct the input image. Cycle consistency loss \cite{9157662} is defined as,

\begin{equation}
L_{cyc} = \mathbb{E}_{x}[\parallel x-G(G(x,s),\widetilde{s})\parallel_{1}]
\end{equation}

where ${s}$ is the target domain style code and ${\widetilde{s} = E(x)}$ the style embedding for the input image ${x}$.

\textbf{Style Consistency Loss:} To ensure the reference and generated images have closely aligned style characteristics, we employ a style consistency loss to enforce style characteristics by reconstructing the style.

\begin{equation}
    L_{sty} = \mathbb{E}_{x}\left[\parallel s-E(G(x,s))\parallel_1\right]
\end{equation}

where ${E(G(x,s))}$ is the reconstructed style code from the output image ${G(x,s)}$ for input $x$ and target style code $s$. When compared to other models \cite{HUANG2018,9157662}, the main difference here is that we only have one style encoder ${E(.)}$ with a single branch that enforces the Generator to bring style characteristics while reducing the need for itself to be trained.

\textbf{Total Loss:} The final total loss that has to be minimized can be expressed as:

\begin{equation}
    L_{total} = L_{adv}+\lambda_{cyc}L_{cyc}+\lambda_{sty}L_{sty}
\end{equation}

where ${\lambda_{cyc},\lambda_{sty}}$ are hyperparameters and are equal to 1.

\section{Experiments and Results}
In this section, the dataset preparation is described, and we analyze the performance of our model with standard baselines CycleGAN \cite{8237506} and MUNIT \cite{HUANG2018}. All the comparative experiments were conducted using the provided author implementations.

\subsection{Dataset Description}
% In this work, we create results based on two publicly available datasets Kermany and rare image dataset. Kermany dataset contains 3 prevalent diseases CNV, DME, Drusen. The rare image dataset contains CSC,MH,Mactel,RP and Stargart that are considered to be rare diseases. Kermany dataset was collected from expert opthamologists where as rare image dataset was prepared by taking images from web.

We use two publicly available datasets provided by \textit{Kermany D}\footnote{https://data.mendeley.com/datasets/rscbjbr9sj/3} and \textit{TaeKeun Yoo}\footnote{https://data.mendeley.com/datasets/btv6yrdbmv/2} to create the results in this work. \textit{Kermany}'s dataset consists of 4 prevalent classes of retinal OCT images: Normal, Drusen \cite{pmid22559899}, DME \cite{pmid22585044}, Chorodial Neovascularization(CNV) \cite{FARIDI2017294}. \textit{Taekeun}'s dataset has 5 diseases that are considered to be rare: central serous chorioretinopathy (CSC), macular hole (MH), retinitis pigmentosa (RP), macular telangiectasia (Mactel) and Stargardt disease. We aim to study the performance of various models given the limited amount of available data.

While \textit{Kermany}'s dataset is a large scale dataset consisting of 27110 normal, 37455 CNV, 11598 DME, and 8866 drusen retinal images, \textit{Taekeun}'s dataset is collected from google with only 30 CSC, 30 MH, 24 Mactel, 19 RP, and 16 Stargardt disease images. For each class, we sample 1000 train images and 100 test images from the \textit{Kermany}'s dataset. For \textit{Taekun}'s dataset we randomly augment the images by shifting from ${-5\%}$ to ${+5\%}$, rotating between ${-15^{\circ}}$ and ${+15^{\circ}}$, scaling up to ${20\%}$, altering brightness between ${-10\%}$ and ${+10\%}$, and elastic transformation \cite{pmid33492598}. We generated 400 images for training and 80 test images for each class. The normal images were utilized in the same ratio taken from \textit{Kermany's} images for training the \textit{Taekun's} dataset as well.

\begin{figure}
  \centering
  \includegraphics[scale=0.6]{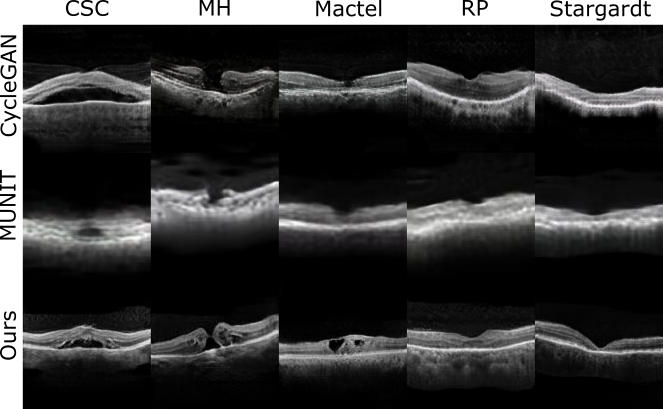}
  \caption{Qualitative comparison of the models on the \textit{Taekun} dataset. Each row corresponds to different models and the columns represent the generated target pathology. For MUNIT and our model, we generate using reference images.}
  \label{fig:fig2}
\end{figure}
\begin{figure}
  \centering
  \includegraphics[scale=0.6]{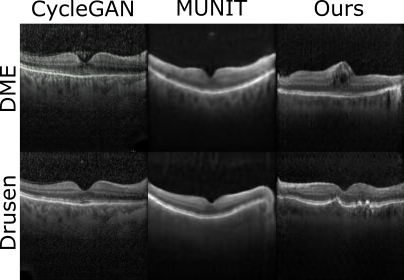}
  \caption{Qualitative comparison of the models on the \textit{Kermany} dataset. Columns and rows represent the model and its generated pathology.}
  \label{fig:fig3}
\end{figure}

\subsection{Experimental Setup and Results}

All of the experiments are done on a single 12GB Nvidia Tesla K80 graphic card with 64GB RAM and Intel Xeon E5-2670 processor. We train the models at $128 \times 128$ resolution with batch size 8 and learning rate ${0.0001}$ for 100 epochs using Adam optimizer \cite{2014arXiv1412.6980K}. For CycleGAN and MUNIT we train multiple models for each normal and disease pair since they can translate between two domains only. To assess style-based translation fairly between MUNIT and our model we evaluate reference-based translation only.

\begin{figure}
  \centering
  \includegraphics[scale=0.6]{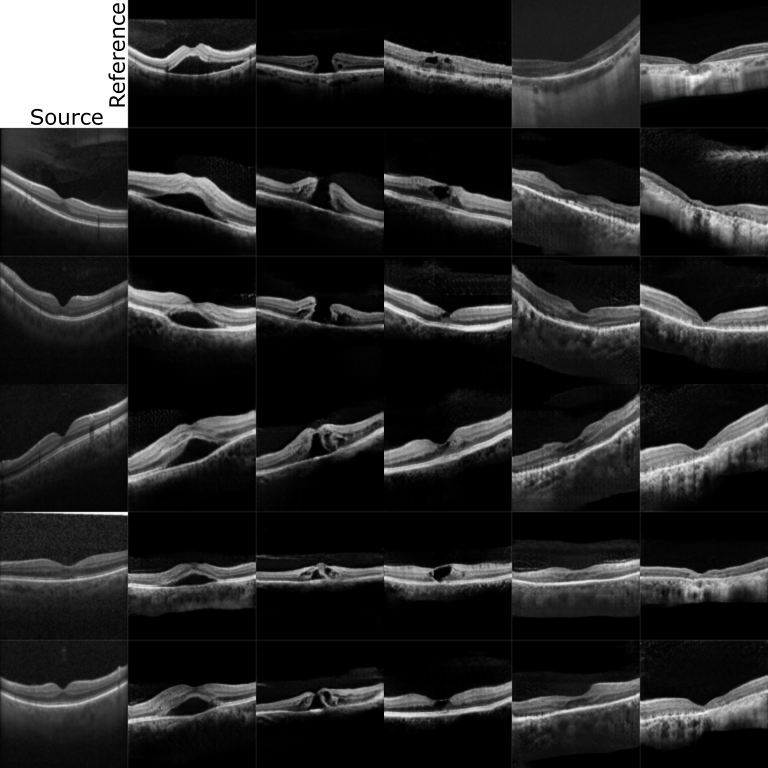}
  \caption{\textbf{Reference based synthesis on \textit{\textbf{Taekun}} dataset.} The first column presents the normal images that are given to our model as the source images while the first row corresponds to various pathologies that are provided as reference images. All the other images are outputs generated by our model translating from normal b-scan to pathological b-scan. Images in each row represent generated outputs for the same source image with different styles or domains. It can be noted that the source domain characteristics are well preserved while translating into the target domain.}
  \label{fig:fig4}
\end{figure}

\textbf{Qualitative evaluation:} Figure \ref{fig:fig2} and \ref{fig:fig3} compares the generated images by the three models for the considered two datasets. We can observe that in Figure \ref{fig:fig2} both CycleGAN and our model generate good quality images while MUNIT is still learning on \textit{Taekun} dataset. The style encoder in the MUNIT needs to be trained along with the generator model which makes the training complex and slow, which is overcome by our pre-trained style encoder thus making the convergence faster. And in Figure \ref{fig:fig3} for \textit{Kermany} dataset, our pre-trained style encoder shows its ability to capture representations that are not prevalent and generate pathologies, where both CycleGAN and MUNIT fail to do so. Figure \ref{fig:fig4} shows the generation of pathological images for various input and reference images. We can see that the proposed model generates the reference pathology while preserving the content characteristics of the input image.

\textbf{Quantitative evaluation:} We report FID \cite{NIPS2017_8a1d6947} and LPIPS \cite{8578166} metrics to evaluate the models both on their quality and diversity. We generate 10 pathological images from each normal retinal image for individual classes and calculate the metrics. Table \ref{tab:table1} shows the obtained metrics for all the models. Since CycleGAN is limited in diversity we don't calculate LPIPS scores for it. We can observe that for \textit{Kermany's} dataset even with the low FID scores, both the CycleGAN and MUNIT fail to generate pathologies. Our model surpasses the performance of both of the models by generating higher quality images while also showing good diversity.

\begin{table}
  \centering
  \begin{tabular}{c|cc|cc}
    \toprule
    &\multicolumn{2}{c}{\textit{Taekun}} & \multicolumn{2}{c}{\textit{Kermany}}
    \\
    \midrule
    Model     & FID ${\downarrow}$     & LPIPS ${\uparrow}$ & FID ${\downarrow}$ & LPIPS ${\uparrow}$  \\
    \midrule
    CycleGAN & 160.72  & -  & 94.19 & -   \\
    MUNIT     & 182.71 & 0.0765 & 89.175 & 0.01997    \\
    Ours     & 108.30   & 0.1710 & 60.8 & 0.14648 \\
    \midrule
    Ground Truth & 35.94 & - & 52.31 & -  \\
    \bottomrule
  \end{tabular}
  \vspace{2mm}
  \caption{Quantitative comparison of the models with reference-guided synthesis for MUNIT and our model.}
  \label{tab:table1}
\end{table}

% The documentation for \verb+natbib+ may be found at
% \begin{center}
%   \url{http://mirrors.ctan.org/macros/latex/contrib/natbib/natnotes.pdf}
% \end{center}
% Of note is the command \verb+\citet+, which produces citations
% appropriate for use in inline text.  For example,
% \begin{verbatim}
%   \citet{hasselmo} investigated\dots
% \end{verbatim}
% produces
% \begin{quote}
%   Hasselmo, et al.\ (1995) investigated\dots
% \end{quote}

% \begin{center}
%   \url{https://www.ctan.org/pkg/booktabs}
% \end{center}

% \subsection{Figures}
% \lipsum[10] 
% See Figure \ref{fig:fig1}. Here is how you add footnotes. \footnote{Sample of the first footnote.}
% \lipsum[11] 

% \subsection{Tables}
% \lipsum[12]
% See awesome Table~\ref{tab:table}.

% \begin{table}
%  \caption{Sample table title}
%   \centering
%   \begin{tabular}{lll}
%     \toprule
%     \multicolumn{2}{c}{Part}                   \\
%     \cmidrule(r){1-2}
%     Name     & Description     & Size ($\mu$m) \\
%     \midrule
%     Dendrite & Input terminal  & $\sim$100     \\
%     Axon     & Output terminal & $\sim$10      \\
%     Soma     & Cell body       & up to $10^6$  \\
%     \bottomrule
%   \end{tabular}
%   \label{tab:table}
% \end{table}

% \subsection{Lists}
% \begin{itemize}
% \item Lorem ipsum dolor sit amet
% \item consectetur adipiscing elit. 
% \item Aliquam dignissim blandit est, in dictum tortor gravida eget. In ac rutrum magna.
% \end{itemize}

\section{Conclusion and Future Work}
In this work, a generative adversarial network model to generate synthetic retinal OCT data was proposed to address the problem of translating between only two domains. The proposed GAN model presents a pre-training style encoder which results in obtaining a robust style code that helps to achieve better results. We also introduce a new discriminator by combining multi-task discriminator and relative discriminator. The model is then evaluated on two distinct datasets and the results show that it can generate good quality images even with limited data and outperform previous models \cite{8237506,HUANG2018} that are remarkably outstanding. Although our model achieves good results, it mainly depends on the style encoder pre-training. We have observed that bad pre-training may not cluster the style embeddings appropriately which in turn affects the discriminator and lose its ability to distinguish between different classes. This results in degradation of generated images and uncontrolled disease synthesis while drastically affecting the training of the model. In future work, different pre-training methods for style encoder and the maximum number of domains that the model can translate robustly need to be explored.  

%\section{Conclusion}
%We proposed a generative adversarial network that addresses the problem of translating between only two domains. We pre-train the style embeddings and use them to train the model. We evaluate the model on two distinct datasets and show that the model can generate good quality images even with limited data and outperform previous models \cite{8237506,HUANG2018} that are remarkably outstanding.

%\section*{Acknowledgments}
%This was was supported in part by......

%Bibliography
\bibliographystyle{unsrt}  
\bibliography{references}

\end{document}